\documentclass[11pt]{article}
\pdfoutput=1 % if your are submitting a pdflatex (i.e. if you have
   
\usepackage{jheppub} % for details on the use of the package, please
\usepackage{url}
\usepackage{caption}
\usepackage{subcaption}

\usepackage{times}

\usepackage[latin9]{inputenc}
\usepackage{float}
\usepackage{amsmath}
\usepackage{amssymb}
\usepackage{dsfont}
\usepackage{graphicx}
\usepackage{esint}
\usepackage{hyperref}
\usepackage{comment}
\usepackage{color}
\usepackage{microtype}
\usepackage{cleveref}
\usepackage{breakurl}
\usepackage{bbm}

\usepackage{mathtools}
\usepackage{enumerate}
\usepackage{tensor}
\usepackage{mathrsfs}
\usepackage{upgreek}
\usepackage{amsbsy}
\usepackage{xcolor}
\hypersetup{
	colorlinks=true,
	linkcolor=blue,
	filecolor=magenta,      
	urlcolor=cyan,
}
\newcommand{\inp} [1] {\left( #1 \right)}
\newcommand{\inb} [1] {\left\{ #1 \right\}}
\newcommand{\insb} [1] {\left[ #1 \right]}
\newcommand {\qmatrix} [1] {\begin{pmatrix} #1 \end{pmatrix}}

\newcommand{\be}{\begin{equation}}
\newcommand{\ee}{\end{equation}}
\newcommand{\ben}{\begin{displaymath}}
\newcommand{\een}{\end{displaymath}}
\newcommand{\bea}{\begin{eqnarray}}
\newcommand{\eea}{\end{eqnarray}}

   \newcommand{\rf}[1]{(\ref{#1})}

\def\be{\begin{equation}}
\def\ee{\end{equation}}
\def\bea{\begin{eqnarray}}
\def\eea{\end{eqnarray}}
\def\bit{\begin{itemize}}
\def\eit{\end{itemize}}

\def\Tr{{\rm Tr}}

\def\R{\mathbb{R}}
\def\Z{\mathbb{Z}}
\def\S{\mathbb{S}}

 \makeatletter
\usepackage{mathtools}
\usepackage{enumerate}
\usepackage{tensor}
\usepackage{mathrsfs}
\usepackage{upgreek}
\usepackage{amsbsy}
\usepackage{xcolor}
\hypersetup{
	colorlinks=true,
	linkcolor=blue,
	filecolor=magenta,      
	urlcolor=cyan,
}
\def\R{\mathbb{R}}
\def\Z{\mathbb{Z}}
\def\S{\mathbb{S}}

\allowdisplaybreaks

\makeatother

\makeatletter
\DeclareRobustCommand{\rcite}[1]{%
  \rcite@aux#1,\@nil{#1}%
}
\def\rcite@aux#1,#2\@nil#3{%
  \if\relax#2\relax
    Ref.~\cite{#3}%
  \else
    Refs.~\cite{#3}%
  \fi
}
\makeatother

\hypersetup{
    colorlinks = true,
    citecolor = {blue},
    linkcolor = {blue},
    urlcolor = {blue},
}

\title{\rm {\LARGE \bf Quantization of Gravity in the Black Hole Background}}

\author{Renata Kallosh and Adel A. Rahman}

\affiliation{Stanford Institute for Theoretical Physics and Department of Physics,\\ Stanford University, Stanford, CA 94305, USA}

\emailAdd{kallosh@stanford.edu}
\emailAdd{arahman@stanford.edu}

\notoc

\abstract{We perform a covariant (Lagrangian) quantization of perturbative gravity in the background of a Schwarzschild black hole. The key tool is a decomposition of the field into spherical harmonics. We fix Regge-Wheeler gauge for modes with angular momentum quantum number $l \geq 2$, while for low multipole modes with $l$ $=$ $0$ or $1$---for which Regge-Wheeler gauge is inapplicable---we propose a set of gauge fixing conditions which are 2D background covariant and perturbatively well-defined. We find that the corresponding Faddeev-Popov ghosts are non-propagating for the $l\geq2$ modes, but are in general nontrivial for the low multipole modes with $l = 0,1$. However, in Schwarzschild coordinates, all time derivatives acting on the ghosts drop from the action and the low multipole ghosts have instantaneous propagators. Up to possible subtleties related to quantizing gravity in a space with a horizon, Faddeev's theorem suggests the possibility of an underlying canonical (Hamiltonian) quantization with a manifestly ghost-free Hilbert space.}

\begin{document}

\maketitle

\newpage

\tableofcontents{}
 
\section{Introduction}

We would like to perform a perturbative quantization of gravity in the background of a Schwarzschild black hole, in a setting originally studied by Regge and Wheeler \cite{Regge:1957td}, and later studied by Zerilli \cite{Zerilli:1971wd} and by Martel and Poisson \cite{Martel:2005ir}, which was recently discussed in the context of quantization by \cite{Gaddam:2020rxb,Gaddam:2020mwe}.  

As is well known, a key obstacle to a straightforward perturbative quantization of gravity is the presence of a gauge symmetry, diffeomorphism invariance, which leaves the na\"ive Feynman path integral ill-defined. A procedure for defining the covariant (Lagrangian) path integral for quantum field theories with gauge symmetries was proposed in \cite{Faddeev:1967fc,DeWitt:1967ub} and involves breaking the gauge symmetry using a set of gauge-fixing conditions and introducing compensating Faddeev-Popov (FP) ghost fields, whose action is computed according to the rules introduced in \cite{Faddeev:1967fc,DeWitt:1967ub}. See \cref{cov} for a review.

In the past such a quantization of gravity  was performed either in a general type of curved background satisfying the Einstein field equations \cite{DeWitt:1967ub,tHooft:1974toh,Kallosh:1974yh,Grisaru:1975ei}, 
with a background-covariant and Lorentz-covariant gauge choice such as background covariant harmonic gauge, or in a flat Minkowski background \cite{Faddeev:1969su,Fradkin:1970pn,Faddeev:1973zb} 
with either a Lorentz-covariant gauge choice such as harmonic gauge or a non-Lorentz-covariant gauge choice such as Dirac gauge
\cite{Dirac:1958jc}.

In the background-covariant and Lorentz-covariant harmonic gauge and its generalizations, the 1- and 2-loop computations in gravity were performed in \cite{tHooft:1974toh,Grisaru:1975ei,Kallosh:1978wt,Barvinsky:1985an,Goroff:1985th,vandeVen:1991gw}. In these gauges the Fadeev-Popov (FP) ghosts are propagating fields which give an important contribution to the Feynman rules. In \cite{tHooft:1974toh,Grisaru:1975ei,Goroff:1985th} the standard background-covariant and Lorentz-covariant de Donder gauge fixing condition 
$F_{\mu} = \nabla_{\nu}h\indices{^\nu_\mu} -\frac{1}{2}\nabla_{\mu}h\indices{^\nu_\nu}$, linear in the metric perturbation $h_{\mu\nu}$, was used. In \cite{vandeVen:1991gw} the choice of the gauge-fixing condition involved in addition to the standard de Donder term a set of extra terms nonlinear in $h_{\mu\nu}$ (for example terms like 
$h^{\nu\lambda} \nabla_\lambda h_{\mu \nu}$),  with arbitrary numerical or depending on scalars of the theory, coefficients. The result, in this case of the 2-loop UV divergence in gravity, is independent of the 7 additional parameters defining the generalized gauge fixing condition. The FP ghost contributions were computed in these generalized classes of gauges according to standard rules,  
and the final results were gauge-independent, since the FP ghosts were taken into account. 

The correct Feynman path integral for computing loop diagrams has been studied both in covariant (Lagrangian) quantization and  canonical (Hamiltonian) quantization, see for example \cite{DeWitt:1967ub,Faddeev:1969su,Fradkin:1970pn,Faddeev:1973zb,Fradkin:1976xa,Fradkin:1977hw,Batalin:1977pb}. We describe the main features of both ways of defining the perturbative Feynman path integral in Appendices \ref{cov}  and \ref{can}, respectively.

Here we discuss briefly the relation between these two types of quantization and their equivalence. In the covariant case, the gauge symmetry of the action $S[\phi] = \int\mathcal{L}(\phi)$ with infinitesimal gauge parameters $\xi_{\alpha}(x)$, $\alpha = 1, \dots, m$ has to be broken in order to lift the degeneracy of the kinetic terms and define the propagators. This is accomplished by adding a gauge-fixing condition to the classical theory, for example in the form  $\chi_\alpha(\phi) =0$, where the functions $\chi_\alpha(\phi)$ depend on the fields $\phi$ of the classical action and possibly their derivatives. The gauge fixing functions $\chi_\alpha$ transform under infinitesimal gauge transformations as 
\be
\delta \chi_\alpha = Q_{\alpha}{}^{\beta} \xi_\beta
\label{gtPrime}
\ee
where $Q_{\alpha}{}^{\beta}$ is in general some differential operator which depends on the fields $\phi$ of the classical action and their derivatives. This then defines the Lagrangian for the Fadeev-Popov (FP) ghosts corresponding to the given gauge-fixing condition via:
\be
{\cal L} _{\mathrm{ghost}}[C,\bar{C}] = \bar C^\alpha  Q_{\alpha}{}^{\beta} C_\beta
\label{gh}\ee where $\bar C^\alpha, C_\beta$ are known as FP anti-ghosts and ghosts, respectively. The gauge-fixing condition $\chi_{\alpha}$ is viewed as admissible provided $\det ||Q_{\alpha}{}^{\beta} || \neq 0$ and provided that arbitrary field configurations $\phi$ can always be transformed to configurations satisfying $\chi_{\alpha}(\phi) = 0$ via gauge transformations respecting the boundary conditions (see \cref{bcs}). Covariant quantization has relatively simple rules which guarantee the gauge-independence of the physical observables in the theory.

One important feature of the covariant quantization is that the local measure of integration is not strictly defined. We refer to \cite{Fradkin:1977hw} where this issue is discussed in detail. In particular, in the case that $Q_{\alpha} {}^{ \beta}(\phi) $ is a local function of fields without differential operators acting on $C_\beta(x)$, the relevant ghost action simply becomes $\bar C^\alpha(x)  \tilde C_\alpha(x)$ where $\tilde C_\alpha(x) \equiv Q_{\alpha} {}^{ \beta} (\phi(x) )C_\beta(x)$. The corresponding ghosts are non-propagating and drop from Feynman rules. The functional determinant resulting from the transformation from $C$ to $\tilde C$ may contribute to the local measure of integration some field dependent divergent term  proportional  to $\delta^D(0)$ where $D$ is the dimension of the space-time on which $x$ is a coordinate (in our case we will have $D = 2$ since our fields will become effectively two-dimensional after decomposition into spherical harmonics). Terms  of this nature can be neglected in a regularized theory or can be shown to cancel, as explained in  \cite{Fradkin:1977hw}.
 
The canonical quantization procedure is more fundamental, being based on a canonical Hamiltonian, but it is also more involved. The issue of unitarity of the S-matrix can only be addressed in the canonical quantization method. However, according to  \cite{Faddeev:1969su,Faddeev:1973zb,Fradkin:1976xa,Fradkin:1977hw,Batalin:1977pb} the result of covariant quantization is equivalent to the canonical one. In the canonical formulation of theories with gauge symmetries, one encounters first class constraints $\phi^\alpha(p_i,q^i)=0$ where $i = 1, \dots, n$ runs over the na\"ive configuration space and again $\alpha=1, \dots , m$. An additional set of conditions $\chi_\alpha(p,q)=0$ has to be added to perform the canonical quantization, and it is required that the Poisson brackets of constraints with the additional conditions have a non-vanishing determinant, $\det ||\{ \chi_\alpha, \phi^\beta\} || \neq 0$. 

 There are two different classes of conditions $\chi_\alpha$. One corresponds to the case of ``unitary" Hamiltonians in gauge theories, which have manifestly ghost-free underlying Hilbert spaces; the other corresponds to the case of ``pseudo-unitary" Hamiltonians in gauge theories, which have underlying state spaces with negative-norm ghost degrees of freedom which must be quotiented out to yield the physical Hilbert space. In the second case the S-matrix is pseudo-unitary in a space of states with the indefinite metric. More details on this are in Sec. \ref{H}.
 
We will show here that in the Regge-Wheeler gauge \cite{Regge:1957td,Zerilli:1971wd,Martel:2005ir} for gravitational perturbations with angular momentum quantum number $l \geq 2$, the corresponding FP ghosts are non-propagating and decoupled in the covariant quantization method. In the low multipole $l=1,0$ sector in the 
two-dimensionally background covariant gauges we propose here, the FP ghosts are present and in general propagating in covariant quantization. However, in Schwarzschild coordinates in  covariant quantization the FP ghosts have ``instantaneous" propagators containing only space derivatives, which suggests that  our covariant quantization may have a corresponding canonical quantization which belongs to the first class of theories mentioned above, with a unitary Hamiltonian and manifestly ghost-free Hilbert space.

We will provide some evidence that this covariant result is in agreement with one originating from a  canonical quantization, as would be expected according to Fadeev's theorem \cite{Faddeev:1969su}. There may be subtleties in directly applying the logic of \cite{Faddeev:1969su} to the situation here, since Schwarzschild coordinates are singular on the event horizon, which was absent in quantization theorems of \cite{Faddeev:1969su,Faddeev:1973zb,Fradkin:1976xa,Fradkin:1977hw}. We leave the study and resolution of such subtleties for future work.

\section{Gravity in the Schwarzschild Black Hole Background}
\quad Consider a 4D asymptotically flat spacetime $(\mathcal{M}, \bar{g}_{\mu\nu})$ whose spacetime manifold $\mathcal{M} = \mathcal{M}_2\times\mathbb{S}^2$ can be endowed with coordinates $(x^{a},\theta^A)$, $a=1,2$ in which the metric takes the form
\begin{equation}
	\bar{g}_{\mu\nu} = g_{\mu\nu} + h_{\mu\nu}
\end{equation}  
with $g_{\mu\nu}$ the metric of the Schwarzschild black hole written in a ``spherically symmetric" form
\begin{equation}
	g_{\mu\nu}\mathrm{d}x^{\mu}\mathrm{d}x^{\nu} = g_{ab}\,\mathrm{d}x^a\mathrm{d}x^b + r^2(x)\,\mathrm{d}\Omega^2_2
	\label{SchldBackgroundGauge}
\end{equation} 
Here $r(x)$ is defined globally and invariantly by 
\begin{equation}
	4\pi r(x)^2 \equiv \mathrm{Area}\Big(\text{$\mathrm{SO}(3)$ orbit of $x$ in $\mathcal{M}$}\Big)
\end{equation}
As noted in \cref{cov}, we work in the context of 4D asymptotically flat general relativity, for which local diffeomorphisms (i.e. diffeomorphisms obeying certain fall-off conditions, see \cref{bcs} and \cref{TrivialFallOffalm,TrivialFallOffAlm}) are gauge symmetries which act infinitesimally as 
\begin{equation}
	\delta\bar{g}_{\mu\nu} = \overline{\nabla}^{(\bar{g})}_{\mu}\xi_{\nu} + \overline{\nabla}^{(\bar{g})}_{\nu}\xi_{\mu}
	\label{GaugeSymmetry}
\end{equation}
where $\overline{\nabla}^{(\bar{g})}_{\mu}$ denotes the (torsion-free) covariant derivative of $\bar{g}_{\mu\nu}$ and where we take the gauge parameter to be the covector\footnote{The corresponding finite gauge transformation is the diffeomorphism generated by $\bar{g}^{\mu\nu}\xi_{\nu}$.} $\xi_{\mu}$. In the context of the background field method within which we will work, the diffeomorphism \cref{GaugeSymmetry} acts on the background $g_{\mu\nu}$ and perturbation $h_{\mu\nu}$ as 
\begin{equation}
	\delta g_{\mu\nu} = 0, \qquad \delta h_{\mu\nu} = \overline{\nabla}^{(\bar{g})}_{\mu}\xi_{\nu} + \overline{\nabla}^{(\bar{g})}_{\nu}\xi_{\mu}
	\label{GaugeSymmetryBF}
\end{equation}
respectively. Note that in this context the diffeomorphisms \cref{GaugeSymmetry} are kept distinct from diffeomorphisms/coordinate transformations of the background $(\mathcal{M}, g_{\mu\nu})$.

\quad Our formalism (based on that of \cite{Martel:2005ir}) will be covariant with respect to two-dimensional background diffeomorphisms/coordinate transformations of $(\mathcal{M}_2, g_{ab})$, i.e. diffeomorphisms/coordinate transformations of the background $(\mathcal{M}, g_{\mu\nu})$ which are constant on the two-sphere. Letting $\mathcal{D}_{a}$ and $\epsilon_{ab}$ denote the covariant derivative and volume form of $(\mathcal{M}_2,g_{ab})$ respectively, it is helpful to define 
\begin{equation}
	r_{a}(x) = \mathcal{D}_{a}r(x) \quad\text{and}\quad t^{a}(x) = -\epsilon^{ab}r_{b}(x)
	\label{rdef}
\end{equation}
which are orthogonal (but not orthonormal) by construction. Here and below all indices will be raised and lowered with the 2D background metric $g_{ab}$. Here $t^a$ is a pseudo-vector which agrees with the stationary Killing vector 
of the Schwarzschild background up to its transformation properties under parity $\epsilon_{ab} \to -\epsilon_{ab}$. Defining 
\begin{equation}
	f(r) \coloneqq g_{ab}r^{a}r^{b} = -g_{ab}t^{a}t^{b} = 1-\frac{2GM}{r}
\end{equation}
we have that
\begin{equation}
	g^{ab} = \frac{1}{f(r)}\inp{-t^{a}t^{b} + r^{a}r^{b}}
\end{equation}

Due to the warp factor $r^2(x)$, there are ``cross-term" Christoffel symbols for the coordinate systems \cref{SchldBackgroundGauge} which prevent the simple factorization
\begin{equation}
	\mathrm{d}x^{\mu}\,\nabla_{\mu} \neq \mathrm{d}x^{a}\mathcal{D}_{a} + \mathrm{d}\theta^{A}D_{A}
\end{equation}
where we have let $\nabla_{\mu}$ denote the covariant derivative of $(\mathcal{M},g_{\mu\nu})$ and $D_A$ denote the covariant derivative of $(\mathbb{S}^2, \Omega_{AB})$. These ``cross-term" Christoffel symbols are given by
\begin{equation}
	\Gamma\indices{^{a}_{A}_{B}} = -r\,g^{ab}r_{b }\,\Omega_{AB}, \qquad \Gamma\indices{^{\mu}_{a}_{B}} = \frac{1}{r}\,r_{a}\,\delta^{\mu}_B
\end{equation}
The remaining ``cross-term" Christoffel symbols vanish, i.e. $\Gamma\indices{^{A}_{a}_{b}} = 0$.

 ``Schwarzschild coordinates"
\begin{equation}
	g_{ab}\,\mathrm{d}x^a\mathrm{d}x^b = -f(r)\,\mathrm{d}t^2 + \frac{\mathrm{d}r^2}{f(r)}
\end{equation}
which will serve an important role in the latter part of this paper, cover the right static patch (right outer domain of communications) of the Schwarzschild black hole and are adapted to its $\Z_2\rtimes \R$ ``static $\rtimes$ stationary" isometry.  Here the ``radial coordinate" $r$ is the function defined by \cref{rdef} and $t$ is an affine parameter along the timelike orbits of $t^a$ which obeys $t^a\mathcal{D}_{a}t = 1$ in the given patch. It is useful to note that, in these coordinates,
\begin{equation}
	r_a\mathrm{d}x^a = \mathrm{d}r, \qquad r^a\partial_a = f(r)\partial_r, \qquad t^a\partial_a = \partial_t
\end{equation}
Note that $t^a\partial_a = \partial_t$ is timelike throughout this patch, asymptoting to a null (pseudo-)vector along the event horizon and a unit $t$ translation at infinity. Note that the slices $\{\Sigma_t\}$ of constant $t$ are everywhere spacelike and constitute a family of Cauchy surfaces for this patch.

\section{Gravitational Perturbations in the Spherical Harmonic Basis}
\subsection{Decomposition into Spherical Harmonics}
It is helpful to take advantage of the spherical isometries and orientability of the background---and, in particular, of the explicit spherical symmmetry of our background gauge choice \cref{SchldBackgroundGauge}---to expand the metric perturbation $h_{\mu\nu}$ in spherical harmonics of definite parity. In the ``spherical" coordinates \cref{SchldBackgroundGauge}, the components of $h_{\mu\nu}$ transform under the action of the background $\mathrm{SO}(3)$ spherical symmetry associated to each two-sphere of fixed $x^{a}$ (and under parity $\epsilon \to - \epsilon$) as three scalars $h_{ab}$, two vectors $h_{a A}$, and one (symmetric) second-order tensor $h_{AB}$. We can thus resolve $h_{\mu\nu}$ into spherical harmonics of definite parity via \cite{Regge:1957td,Zerilli:1971wd,Martel:2005ir}
\begin{equation}
	h_{\mu\nu} = \underbrace{\qmatrix{p_{ab} & p^{(+)}_{a B} \vspace{0.5em}\\
			p^{(+)}_{b A} & p^{(+)}_{AB}}}_{\text{parity even}} \ + \ \underbrace{\qmatrix{0 & p^{(-)}_{a B} \vspace{0.5em}\\ p^{(-)}_{Ab } & p^{(-)}_{AB}}}_{\text{parity odd}}
	\label{hdecomp}
\end{equation}
with
\begin{align}
	p_{ab} &= \sum_{\ell = 0}^{\infty}\sum_{|m|\leq \ell}h_{ab}^{\ell m}Y^{\ell m} & &
	\label{SHBHScalar}\\
	p_{a A}^{(+)} &= \sum_{\ell = 1}^{\infty}\sum_{|m|\leq \ell}j_{a}^{\ell m}\,Y^{\ell m}_{A} & p_{a A}^{(-)} &= \sum_{\ell = 1}^{\infty}\sum_{|m|\leq \ell}h_{a}^{\ell m}\,X^{\ell m}_{A} \label{SHBHVector}\\
	p^{(+)}_{AB} &=  r^2\inp{\sum_{\ell = 0}^{\infty}\sum_{|m|\leq \ell}K^{\ell m}\Omega_{AB}Y^{\ell m} + \sum_{\ell = 2}^{\infty}\sum_{|m|\leq \ell}G^{\ell m}Y^{\ell m}_{AB}} & p^{(-)}_{AB} &= \sum_{\ell = 2}^{\infty}\sum_{|m|\leq \ell}h_2^{\ell m}X^{\ell m}_{AB} \label{SHBHTensor}
\end{align}

Here $Y^{\ell m}(\theta^A)$ are the usual unit-normalized spherical harmonics on $\mathbb{S}^2$. The even-parity vector and tensor spherical harmonics are defined by ${Y^{\ell m}_A \equiv D_AY^{\ell m}}$ and ${Y^{\ell m}_{AB} \equiv \insb{D_AD_B + \frac{1}{2}\ell\inp{\ell + 1}\Omega_{AB}}Y^{\ell m}}$ respectively. The odd-parity vector and tensor spherical harmonics are defined by ${X^{\ell m}_A \equiv -\epsilon\indices{_A^B}D_BY^{\ell m}}$ and ${X^{\ell m}_{AB} \equiv -\frac{1}{2}\insb{\epsilon\indices{_A^C}D_CD_B + \epsilon\indices{_B^C}D_CD_A}Y^{\ell m}}$ respectively, where $\epsilon_{AB}$ is the volume form of the round unit sphere and all indices have been raised and lowered with $\Omega_{AB}$. 

Note that $p_{ab}$ can be regarded as an $\mathcal{M}_2$ tensor and an $\mathbb{S}^2$ scalar; $p_{aA}$ as both an $\mathcal{M}_2$ vector and an $\mathbb{S}^2$ vector; and $p_{AB}$ as an $\mathcal{M}_2$ scalar and an $\mathbb{S}^2$ tensor. The decomposition \cref{hdecomp,SHBHScalar,SHBHVector,SHBHTensor} completely specifies the angular dependence of the perturbation $h_{\mu\nu}$. In particular, the coefficient functions $h_{ab}^{\ell m}$, $j_{a}^{\ell m}$, $K^{\ell m}$, $G^{\ell m}$, $h_{a}^{\ell m}$, $h_2^{\ell m}$ are ``angle independent" (constant on the two-sphere and hence functions only of the two-coordinate $x^a$) and can be regarded as scalar, vector and tensor fields on $\mathcal{M}_2$.

\subsection{Gauge Freedom in Spherical Harmonics} 
Note that, since the different spherical harmonics decouple from one another in the kinetic term of the action, we can make independent gauge choices  
for each choice of $(\pm)\ell m$. In order to understand how the gauge freedom \cref{GaugeSymmetryBF}
\begin{equation}
	h_{\mu\nu} \to h_{\mu\nu} + \overline{\nabla}^{(\bar{g})}_{\mu}\xi_{\nu} + \overline{\nabla}^{(\bar{g})}_{\nu}\xi_{\mu}
	\label{GaugeInv}
\end{equation}
acts on a given harmonic, we resolve the gauge parameter $\xi_{\mu}$ into spherical harmonics as
\begin{equation}
	\xi_{\mu} = \underbrace{(\Xi_a,\Xi_A^{(+)})}_{\text{parity even}} \ + \ \underbrace{(0,\Xi_A^{(-)})}_{\text{parity odd}}
\end{equation}
with
\begin{equation}
	\Xi_{a} = \sum_{\ell = 0}^{\infty}\sum_{|m|\leq \ell}\xi_{a}^{\ell m}Y^{\ell m}
	\label{DiffHarmonicsScalar}
\end{equation}
\begin{equation}
	\Xi_A^{(+)} = \sum_{\ell = 1}^{\infty}\sum_{|m|\leq \ell}\xi^{(+)\ell m}Y^{\ell m}_A, \qquad \Xi_A^{(-)} = \sum_{\ell = 1}^{\infty}\sum_{|m|\leq \ell}\xi^{(-)\ell m}X^{\ell m}_A
	\label{DiffHarmonicsVector}
\end{equation}
The $\xi_{a}^{\ell m}$, $\xi^{(+)\ell m}$, $\xi^{(-)\ell m}$ are ``angle independent" (constant on the two-sphere and hence functions only of the two-coordinate $x^a$) and 
can be regarded as scalar and vector fields on $\mathcal{M}_2$.

\

In terms of these variables the gauge transformation \cref{GaugeSymmetryBF} reads
 \begin{align}
 	\delta p_{ab} &= \mathcal{D}_a\Xi_b + \mathcal{D}_b\Xi_a -2\widehat{\Gamma}^{\mu}_{ab}\xi_{\mu}\cr
 	\delta p_{aB} &= \mathcal{D}_a\Xi_B + D_B\Xi_a-\frac{2}{r}\,r_a\Xi_b-2\widehat{\Gamma}^{\mu}_{aB}\xi_{\mu}\cr
 	\delta p_{AB} &= D_A\Xi_B + D_B\Xi_A + 2rg^{ab}r_a\Xi_b\Omega_{AB}-2\widehat{\Gamma}^{\mu}_{AB}\xi_{\mu}
\label{gt}
\end{align} 
The last term involves the linear tensor $\widehat{\Gamma}^{\rho}_{\mu\nu}$ which relates the full $\overline{\nabla}_{\mu}^{(\bar{g})}$ and background $\nabla^{(g)}_{\mu}$ covariant derivatives: 
\begin{equation}
	\widehat{\Gamma}^{\rho}_{\mu\nu}\xi_{\rho} \equiv \overline{\nabla}_{\mu}^{(\bar{g})}\xi_{\nu} - \nabla^{(g)}_{\mu}\xi_{\nu}
	\label{ChristoffelTerms}
\end{equation}
\begin{equation}
	\widehat{\Gamma}^{\rho}_{\mu\nu} \equiv \frac{1}{2}\,\bar{g}^{\rho\sigma}\Big ( {\nabla^{(g)}_{\mu}h_{\sigma\nu} + \nabla^{(g)}_{\nu}h_{\mu\sigma}-\nabla^{(g)}_{\sigma}h_{\mu\nu}}\Big)
	\label{ChrisExplicit}
\end{equation}

The key fact that will be useful to us is that such terms do not involve derivatives acting on the gauge parameters $\xi_\mu = (\Xi_a, \Xi ^{(\pm)})$, and the contributions of such terms to any FP ghost actions will therefore not involve derivatives acting on ghosts. Since any such terms are at least linear in $h_{\mu\nu}$, they do not contribute to the ghost propagators, but rather provide the couplings for interaction vertices of the form $\bar{C}h^nC$ with $n \geq 1$. It is important to stress here that such terms only involve objects which, after decomposing into spherical harmonics, have coefficients which are proper scalars, vectors, or tensors on $\mathcal{M}_2$, which can be understood after close inspection of \cref{ChrisExplicit}.

In general, the terms \cref{ChristoffelTerms} are complicated power series in $h_{\mu\nu}$ which couple different harmonics of the expansions \cref{SHBHScalar,SHBHVector,SHBHTensor,DiffHarmonicsScalar,DiffHarmonicsVector}, and outside of the monopole ($l = 0$) sector we do not have explicit closed-form expressions for the sums of these series. The general properties of these terms that we will use are that 1) they do not contribute to ghost propagators and 2) their decomposition into harmonics/2D fields only involves proper representations of the Lorentz group (in particular, the coefficients of all terms in the decomposition and expansion will be proper Lorentz scalars). This latter fact will be used in Appendix \ref{nl} where we study the beyond leading order in $h_{\mu\nu}$ part of the $l \geq 1$ ghost actions. These general properties will be sufficient to establish our results. In the monopole ($l = 0$) sector, we have been able to find an explicit closed-form expression \cref{MonopoleExplicit}, valid to all orders in perturbation theory which explicitly displays these two properties. 

\subsection{Boundary Conditions}
\label{bcs}
In gravity, the issue of boundary conditions is especially important, since this will affect the asymptotics of the spacetime on top of which our fields propagate, as well as important physical quantities such as the ADM mass and the flux of gravitational radiation measured at infinity. In the context of asymptotically flat gravity that we consider here, this additional consideration places a restriction on the asymptotic falloff of the gauge parameters $\xi_{\mu}$ which relate physically equivalent field configurations of $h_{\mu\nu}$.  All other choices of $\xi_{\mu}$ correspond to ``large" diffeomorphisms which act nontrivially on the physical space of states\footnote{An example of a family of such ``large" diffeomorphisms which act nontrivially on the physical space of states of asymptotically flat gravity are the BMS Supertranslations \cite{BBM,Sachs}. See e.g. \cite{Strominger} for a recent review.}. 

For 4D general relativity with asymptotically flat boundary conditions\footnote{As noted in \cite{SatishchandranWald}, these asymptotics are slightly weaker than requiring smoothness at future null infinity $\mathscr{I}^+$.}
\cite{ADM,Faddeev:1973zb,SatishchandranWald}
\begin{equation}
h_{\mu\nu} \ \underset{r\to\infty}{\sim} \ O\bigg(\frac{1}{r}\bigg), \qquad \partial_{\rho}h_{\mu\nu} \ \underset{r\to\infty}{\sim} \ O\bigg(\frac{1}{r^2}\bigg)
\label{AFFallOff}
\end{equation} 
the gauge parameters which relate physically equivalent field configurations of $h_{\mu\nu}$ are those which fall off at least as fast as
\begin{equation}
	\bar{g}^{\mu\nu}\xi_{\nu} \ \underset{r\to\infty}{\sim} \ O\bigg(\frac{1}{r}\bigg), \qquad \bar{g}^{\mu\nu}\partial_{\rho}\xi_{\nu} \ \underset{r\to\infty}{\sim} \ O\bigg(\frac{1}{r^2}\bigg)
	\label{TrivialFallOff}
\end{equation} 
or, in terms of the decomposition \cref{DiffHarmonicsScalar,DiffHarmonicsVector},
\begin{equation}
	\xi_a^{\ell m} \ \underset{r\to\infty}{\sim} \ O\bigg(\frac{1}{r}\bigg), \qquad \partial_{r}\xi_a^{\ell m} \ \underset{r\to\infty}{\sim} \ O\bigg(\frac{1}{r^2}\bigg)
	\label{TrivialFallOffalm}
\end{equation}
and 
\begin{equation}
	\frac{1}{r^2}\,\xi^{(\pm)\ell m} \ \underset{r\to\infty}{\sim} \ O\bigg(\frac{1}{r}\bigg), \qquad \frac{1}{r^2}\,\partial_{r}\xi^{(\pm)\ell m} \ \underset{r\to\infty}{\sim} \ O\bigg(\frac{1}{r^2}\bigg)
	\label{TrivialFallOffAlm}
\end{equation} 
In \cref{AFFallOff,TrivialFallOff,TrivialFallOffalm,TrivialFallOffAlm}
we work with the coordinates and radial function \cref{rdef} of the background \cref{SchldBackgroundGauge}, but these statements hold more generally for any asymptotically Minkowskian coordinate system with $r = \sqrt{\|\vec{x}\|^2}$ its spatial radial function. \cref{TrivialFallOff} can be easily proven using the covariant phase space methods of \cite{WaldLee,WaldIyer}. The essential point is that an infinitesimal diffeomorphism of an asymptotically flat metric represents a zero mode of the (appropriately restricted) presymplectic form associated to the 4D Einstein-Hilbert action (this is the precise sense in which $\xi_{\mu}$ relates physically equivalent field configurations) if and only if certain surface integrals at spacelike infinity $i_0$ vanish. With the asymptotically flat boundary conditions \cref{AFFallOff}, this requires the falloff conditions \cref{TrivialFallOff}.

It is important that our gauge-fixing conditions $\chi_{\alpha}$ be chosen such that any field configuration of $h_{\mu\nu}$ can always be put into a form satisfying $\chi_{\alpha} = 0$ via a gauge transformation whose generator satisfies \cref{TrivialFallOffalm,TrivialFallOffAlm}. This will ensure that the corresponding gauge slice includes (at least) one representative from each gauge equivalence class of field configurations. We have been careful to check that this is the case for each of the gauge-fixing conditions chosen below.

The falloff conditions \cref{TrivialFallOffalm,TrivialFallOffAlm}  exclude transformations of the form \rf{GaugeInv} for which the gauge parameters $\xi_\mu$ do not vanish at infinity. This clarifies an issue raised in \cite{Martel:2005ir} concerning the gauge-fixing of the low multipole $l = 0,1$ modes. It was observed in \cite{Martel:2005ir} that their choice of gauge-fixing conditions did not seem to fully determine the gauge, since their conditions were preserved by particular families of additional gauge transformations, with particular generators $\xi_{\mu}(v,r)$ presented in ingoing Edington-Finkelstein (EF) coordinates. However, these generators all violate the conditions \cref{TrivialFallOffalm,TrivialFallOffAlm}: rather than falling off at large $r$, they are either constant or grow with $r$.

After imposing the falloff conditions \cref{TrivialFallOffalm,TrivialFallOffAlm}, the gauge-fixing conditions for low multipoles suggested in \cite{Martel:2005ir} in EF coordinates, or the coordinate independent ones suggested here,  fully determine the gauge\footnote{They also already fully determine the gauge beyond linear order in $h_{\mu\nu}$ %They are also fully determined beyond linear order in $h_{\mu\nu}$
}. Note that such falloff conditions are also required in quantum field theories with gauge symmetries in order to derive the Ward identities which govern the physical observables. 

\subsection{Regge-Wheeler-Zerilli-Martel-Poisson Gauge}
Regge-Wheeler gauge \cite{Regge:1957td} (see also Zerilli \cite{Zerilli:1971wd} and Martel-Poisson \cite{Martel:2005ir}) was often used in studies of linearized perturbations of the classical Einstein equations near the Schwarzschild black hole background. We will now review this gauge-fixing condition and its inapplicability for low-multipole modes with $l = 0,1$, in anticipation of its use in covariant quantization in \cref{CovQuant} below.

As noted above (see \cref{hdecomp,SHBHTensor}), after decomposition into spherical harmonics, the gravitational field $h_{\mu\nu}(x^a, \theta^A)$ 
becomes encoded in a set of functions, 
which, as noted above, may be regarded as 2D fields on $\mathcal{M}_2$. In the notation of \cite{Martel:2005ir}, these functions/2D fields are
\be
h_{ab}^{\ell m(+)}, \quad j_{a}^{\ell m(+)}, \quad K^{\ell m(+)}, \quad G^{\ell m(+)}, \quad h_{a}^{\ell m(-)}, \quad h_2^{\ell m(-)}\,  \qquad l\geq 2
\label{ansatz} \ee
At $l\geq 2$ all functions/2D fields in eq. \rf{ansatz} are available, while for low multipoles only a restricted set of fields is available due to a lack of vector and/or tensor harmonics for those modes. One finds for electric dipoles, magnetic dipoles, and monopoles respectively, the following fields
\bea
&&h_{ab}^{1 m(+)}, \quad  j_{a}^{1 m(+)},  \quad K^{1m(+)},  \qquad l=1, \quad  {\rm even}
\label{ansatzDe} \\
\cr
&&h_{a}^{1 m(-)}  \hskip 4.4 cm  l=1, \quad   {\rm odd}\\
\cr
&&h_{ab}^{0 0(+)},  \quad K^{00(+)}, \hskip 2.6 cm  l=0
\label{ansatzM} 
 \eea
All  functions in \rf{ansatz}-\rf{ansatzM}  can be regarded as fields on $\mathcal{M}_2$: $h_{ab}$ may be regarded as a tensor on $\mathcal{M}_2$, $j_a, h_a$ as vectors on $\mathcal{M}_2$, and $K, G, h_2$ as scalars  on $\mathcal{M}_2$. The angular dependence of the perturbation $h_{\mu\nu}$ 
is encoded in the discrete dependence of these fields on the labels $(\pm) l m$. 
  
As also noted above (see \cref{DiffHarmonicsScalar,DiffHarmonicsVector}), after decomposition into spherical harmonics, the gauge parameters $\xi_{\mu}(x^a, \theta^A)$ also become encoded in a set of functions 
which may be regarded as 2D fields on $\mathcal{M}_2$. At $l\geq 2$ for each $l,m$ there are 4 gauge symmetries. At $l=1(+) $ there are 3 gauge symmetries for each $m=-1,0,1$, at $l=1(-) $ there is one gauge symmetry for each $m=-1,0,1$ and at $l=0$ there are two gauge symmetries. 
\bea
\xi^{l\geq 2} \qquad   &&\Rightarrow \qquad \{ \xi_{a}^{\ell m(+)}, \xi^{\ell m(+)}, \xi^{\ell m(-)} \}
\label{sym}\\
\cr
\xi^{l=1(+)}  \qquad   &&\Rightarrow \qquad  \{ \xi_{a}^{1 m(+)}, \xi^{1 m(+)} \} 
\label{symDe} \\
\cr
\xi^{l=1(-)}  \qquad  &&\Rightarrow \qquad  \{  \xi^{1 m(-)}  \} \\
\cr
\xi^{l=0}  \qquad   &&\Rightarrow \qquad  \{  \xi_{a}^{0 0(+)} \}
\label{symM} 
 \eea
All  functions in \rf{sym}-\rf{symM} can similarly be regarded as fields on $\mathcal{M}_2$: $\xi_a$ may be regarded as a vector on $\mathcal{M}_2$ and $\xi^{(\pm)}$ may be regarded as scalars on $\mathcal{M}_2$. The angular dependence of the gauge parameter $\xi_{\mu}$ is similarly encoded in the discrete dependence of these fields on the labels $(\pm) l m$.

The Regge-Wheeler gauge condition for modes with $l\geq 2$ is

\be
 j_{a}^{\ell m(+)}= G^{\ell m(+)}= h_2^{\ell m(-)} =0 
\label{RW} \ee
It involves one vector and two scalars, leading to a total of four gauge-fixing conditions. One can check that an arbitrary configuration of the fields involved in \cref{RW} can always be brought to Regge-Wheeler gauge by means of a gauge transformation whose generator satisfies the fall-off conditions \cref{TrivialFallOffalm,TrivialFallOffAlm}, so these facts together tell us that there is always one and only one field configuration of (the $l \geq 2$ part of) $h_{\mu\nu}$ in each gauge equivalence class which satisfies the Regge-Wheeler gauge condition \cref{RW}. This tells us that Regge-Wheeler gauge is a good gauge-fixing condition for covariant quantization. Note that the Regge-Wheeler gauge condition \cref{RW} is independent of the choice of coordinates $x^a$ on  $\mathcal{M}_2$.

Some of the fields involved in \cref{RW} are absent at $l<2$ as one can see in eqs. \rf{ansatzDe}-\rf{ansatzM}. Therefore Regge-Wheeler gauge is not a valid gauge choice for low multipole modes with $l < 2$, and we will present an alternative choice in \cref{CovQuant} below.

\subsection{Comments on Monopoles $l=0$ and Dipoles $l=1$}
An important feature of the low multipole modes established in \cite{Zerilli:1971wd,Martel:2005ir} is the following.
First off,  the Regge-Wheeler gauge is not valid  and one has to impose a different set of gauges at $l=0,1$ since for these modes some of the functions in eq. \rf{RW} are absent. Examples of such gauge conditions were proposed in \cite{Zerilli:1971wd,Martel:2005ir}, where it was also observed that the classical equations of motion for linearized perturbations in this sector have simple local in time solutions. 

This is associated with the feature of the low multipoles $l = 0$ and $ l= 1$ that they do not contain radiative degrees of freedom.  The gravitational perturbations near future null infinity were studied in \cite{Martel:2005ir} in  the retarded coordinate system $(u, r, \theta, \phi)$ where $u= t-r-2M \ln (r/2M -1)$. It was shown there that the energy  carried away by
the gravitational radiation near future infinity at $u, r \rightarrow \infty$ is proportional to $l(l-1)$. At the event horizon the radiation was studied in advanced coordinates $(v, r, \theta, \phi)$ where $v= t+r-2M \ln (r/2M -1)$ and again the result is proportional to $l(l-1)$. In both cases the radiation involves quadrupoles and higher modes, monopoles and dipoles drop from radiation in agreement with the standard expectation that $l = 0$ and $ l= 1$ perturbations do not contain radiative degrees of freedom.

In absence of additional sources all solutions of Einstein equations for perturbations with $l=0,1$ can be gauged away according to \cite{Zerilli:1971wd,Martel:2005ir} by an appropriate choice of coordinate transformation. In the presence of additional sources, like a point particle of mass $m_0$ moving  towards the black hole, or a particle orbiting a black hole with a fixed angular momentum $a$, solutions for linearized perturbations take a specific form.  In the $l=0$ monopole case one finds  $h_{tt} \sim  {m_0\over r}$, \cite{Zerilli:1971wd}.  This solution provides a linear correction to the black hole mass proportional to $m_0$. In the $l=1$ odd parity (magnetic dipole) case, the perturbed metric is shown to describe a slightly rotating black hole: One finds according to  \cite{Zerilli:1971wd} that the $l=1$ solution for odd-parity perturbations is of  the form $\sim {m_0 a  \over r^2}$ which represents the linearization of the Kerr metric with respect to its angular-momentum parameter, determined here by the angular momentum of the source.
 
In \cite{Martel:2005ir} where advanced time coordinates were used, the monopole solution is $h_{vv} \sim {2\delta M \over r}$ and the magnetic dipole solution is $h_v\sim {2 \delta J \over r}$. This again confirms that classical solutions for small perturbations lead to small changes in the black holes mass (position of the horizon) and add a small rotation. Finally for the $l=1$ even-parity (electric dipole) case, solutions for small perturbations are vanishing, even in presence of sources, and are interpreted as simply encoding a switch to a non-inertial coordinate system with respect to the original Schwarzschild space-time \cite{Martel:2005ir}.
 
\section{Covariant Quantization of Gravity in the Schwarzschild Background}
\label{CovQuant}
\subsection{Gauge-Fixed Action}\label{BRST}

The general form of the BRST-invariant gauge-fixed action \cite{Becchi:1975nq,Tyutin:1975qk} consists of 3 terms: the classical action, the gauge-fixing part of the action and the FP ghost action. In case of gravity it takes the form
\be
S_{\mathrm{gf}}(g, B, \bar C, C;h) = S_{\mathrm{cl}}(g+h) + \int B^\alpha \chi_\alpha(g; h)  + \int \bar C^\alpha Q_\alpha {}^\beta (g; h)C_\beta
\ee
where the gauge-fixing conditions $\chi_\alpha=0$ result from integrating out the auxiliary fields $B^\alpha$.
The Feynman path integral acquires the form shown in eq. \rf{pi}.

We  now propose the following (two-dimensionally) background covariant and perturbatively well-defined gauge-fixing conditions, including the Regge-Wheeler case \rf{RW} as well as a gauge-fixing of the low multipole modes: 
\bea
&&j_{a}= G= h_2 =0  \hskip 1 cm l \geq  2 , \quad  {\rm even},  \quad  {\rm odd}\cr
\cr
&&K= j_a=0   \hskip 1.8 cm  l=1, \quad  {\rm even}\cr
\cr
&&r^a h_a=0  \hskip 2.3 cm  l=1, \quad  {\rm odd}\cr
\cr
&&K = t^ar^bh_{ab} = 0  \hskip 1 cm l = 0, \quad  {\rm even}
\label{ourG} \eea
All gauge fixing functions here as well as in \rf{RW} are 2D scalars or vectors, with exception of the case $t^ar^bh_{ab} = 0$ which is a pseudoscalar. This simply means that all  auxiliary fields $B^\alpha$ for our choice of gauge-fixing functions $\chi_\alpha$ are 2D scalars or vectors, with the exception of the case $ Bt^ar^bh_{ab}$ where $B$ is a pseudoscalar, so that the total $\int B^\alpha \chi_\alpha(g; h)$ contribution to the action is a 2D Lorentz 
scalar, including the monopole term $ Bt^ar^bh_{ab}$. One can check that arbitrary field configurations of $h_{\mu\nu}$ can be made to satisfy eq. \rf{ourG} by acting with gauge transformations whose generators satisfy \cref{TrivialFallOffalm,TrivialFallOffAlm}. Since \rf{ourG} also exhausts the gauge freedom \rf{GaugeInv}, we see that there is always one and only one field configuration of $h_{\mu\nu}$ in each gauge equivalence class which satisfies the gauge condition \rf{ourG}. This tells us that \rf{ourG} is a good gauge-fixing condition for covariant quantization.

\subsection {Covariant Quantization of the $l\geq 2$ Modes: Decoupling of Ghosts}

We apply to our theory the De Witt-Faddeev-Popov procedure \cite{DeWitt:1967ub,Faddeev:1967fc} for the covariant quantization of quantum field theories with gauge symmetries, see Appendix \ref{cov} for a short review. We choose the four Regge-Wheeler gauge-fixing  conditions used in \cite{Regge:1957td,Zerilli:1971wd,Martel:2005ir} for all modes starting with quadrupoles and above ($l \geq 2$),  is  given, for each $(l,m)$, in eq. \rf{RW}. After writing the gauge symmetry \rf{GaugeInv} in terms of the 
spherical harmonic decomposition \cref{hdecomp,SHBHScalar,SHBHVector,SHBHTensor} for the gravitational field, we will show here that all FP ghosts corresponding to the gauge-fixing conditions \rf{RW} for even and odd  $l\geq 2$ modes decouple.

From \cref{gt}, we see that the gauge transformation \cref{gtPrime} of each of the gauge fixing functions in \rf{RW} is:
\bea
\delta j_{a}^{(+)} &= &  \xi_{a}^{(+)} + \mathcal{D}_{a}\xi^{(+)}  - \frac{2}{r}\,r_{a}\xi^{(+)} + f^{(+)}_{a}[h,\xi],     \label{jaFreedomgeq2}\\
\delta G^{(+)} &=& \frac{2}{r^2}\,\xi^{(+)}  + f_{(G)}^{(+)}[h,\xi] ,  \label{GFreedom}\\
\delta h_2^{(-)} &=& 2\xi^{(-)} + f_2 ^{(-)}[h,\xi], \label{h2Freedom}
\eea
Here the last terms in these equations $f[h, \xi]$ are functionals of the background metric, the gauge parameters $\xi_{\mu}$, and the metric perturbation $h_{\mu\nu}$ or spacetime derivatives of $h_{\mu\nu}$ which obey $f[0,\xi] = 0$ and tensorial linearity in $\xi_{\mu}$ (and hence linearity in $\xi_a$ and $\xi^{(\pm)}$, with no dependence on derivatives of $\xi_a$ and $\xi^{(\pm)}$). These terms originate from the tensor \cref{ChristoffelTerms} which relates the covariant derivative of the background to the full covariant derivative appearing in \cref{gt}. As explained above, the contribution to the ghost action due to these terms does not involve derivatives acting on ghosts, and since these terms are at least  linear or higher power in $h$, they do not contribute to the ghosts propagators, but rather define couplings for the interaction terms $\bar C h^n C$ with $n\geq 1$.

The total ghost Lagrangian is given by
\bea
&&\bar C^a \Big (C_{a}^{(+)} + (\mathcal{D}_{a}  - \frac{2}{r}\,r_{a})C^{(+)} + f^{(+)}_{a}[h,C]\Big ) + \bar{C}^{(+)} \Big(  \frac{2}{r^2}\,C^{(+)}  + f_{(G)}^{(+)}[h,C]  \Big)\cr
&&+ \bar C^-  \Big(  2C^{(-)} + f_2 ^{(-)}[h,C] \Big)
\eea
where we have not specified the details of the interaction terms. The odd sector ghost action is algebraic and decoupled from the even sector. We can integrate out $\bar C^{(-)} $ which leads to the constraint $2C^{(-)} + f_2 ^{(-)}[h,C]=0$, so that the odd ghost action vanishes.

We now integrate over $\bar C^{(+)}$ to find the constraint
\be
\frac{2}{r^2}\,C^{(+)}  + f_{(G)}^{(+)}[h,C] =0
\label{fG}\ee
The remaining ghost Lagrangian is
\be
\bar C^a \Big (C_{a}^{(+)} - {r^2\over 2} (\mathcal{D}_{a}  - \frac{2}{r}\,r_{a}) f_{(G)}^{(+)}[h,C] + f^{(+)}_{a}[h,C]\Big ) 
\label{l2}\ee
where we have used eq. \rf{fG}.
To establish the perturbative Feynman rules we
will now look only at the terms quadratic in ghosts and anti-ghosts, without couplings to $h_{\mu\nu}$ (i. e. neglecting terms with $f[h, C]$) to find the ghosts propagators in the Regge-Wheeler gauge in the background of the Schwarzschild black hole. We find that the terms quadratic in quantum fields are just
\be
\bar C^{a(+)} C_{a}^{(+)} 
\ee
Thus, the quadratic term for the remaining ghosts is
algebraic. We conclude that  all ghosts for the Regge-Wheeler gauge  are non-propagating and can therefore be neglected for all even and odd $l\geq 2$ modes. We study the non-linear part of the ghost action in the Appendix \ref{nl} where we argue that in Schwarzschild coordinates there are no time derivatives acting on the ghosts at all in the $l\geq 2$ sector.

\subsection{Covariant Quantization of the $l=1$ Even (Electric Dipole) Modes}
From \cref{gt}, we see that the gauge transformation \cref{gtPrime} of each of the gauge fixing functions in the $l = 1$ even (electric dipole) part of \rf{ourG} is:
\bea
\delta j_{a}^{(+)} &= &  \xi_{a}^{(+)} + \mathcal{D}_{a}\xi^{(+)}  - \frac{2}{r}\,r_{a}\xi^{(+)} + f^{(+)}_{a}[h,\xi],     \label{jaFreedom1}\\
\delta K^{(+)}&= &-\frac{2}{r^2}\,\xi^{(+)} + \frac{2}{r}\,r^{a}\xi^{(+)}_{a} + f_{(K)}^{(+)}[h,\xi]
\label{KFreedom}\eea
The total ghost Lagrangian is therefore given by
\bea
&&\bar C^a \Big (C_{a}^{(+)} + (\mathcal{D}_{a}  - \frac{2}{r}\,r_{a})C^{(+)} + f^{(+)}_{a}[h,C]\Big ) \cr
&&+ \bar{C}^{(+)} \Big(  -\frac{2}{r^2}\,C^{(+)} + \frac{2}{r}\,r^{a}C^{(+)}_{a} + f_{(K)}^{(+)}[h,C]  \Big)
\eea
We integrate over $ \bar{C}^{(+)}$ to yield the constraint
\be
C^{(+)}=   r r^{a}C^{(+)}_{a} +  \frac{r^2}{2}\, f_{(K)}^{(+)}[h,C] 
\ee
The remaining ghost Lagrangian is 
\bea
&&\bar C^a \Big (C_{a}^{(+)} + (\mathcal{D}_{a}  - 2\,r_{a})( r^{a}C^{(+)}_{a} +  \frac{r}{2}\, f_{(K)}^{(+)}[h,C]) + f^{(+)}_{a}[h,C]\Big ) 
\eea
The part of this Lagrangian quadratic in ghosts (which defines their propagator) is now
\be
\bar C^a \Big (C_{a}^{(+)} + (\mathcal{D}_{a}  - 2\,r_{a}) r^{b}C^{(+)}_{b} \Big ) 
\label{leq1Ghost}
\ee
Thus, there is in general a propagating ghost field $C^{(+)}_a$.

In Schwarzschild coordinates ($t,r$) where $r_a =(0,1) , r^a= (0, f)$, \cref{leq1Ghost} becomes
\be
\bar C^t C_{t}^{(+)} + \bar C^r C_{r}^{(+)} + \bar C^t \mathcal{D}_{t} f C_r +\bar C^r (\mathcal{D}_{r}
- 2\,) f C^{(+)}_{r}
\ee
This action is linear in $C_{t}^{(+)}$, which we can therefore integrate out to impose the constraint $\bar C^t=0$. The remaining part of the Lagrangian quadratic in the ghosts is
\be
\bar C^r [ (\mathcal{D}_{r}
- 2\,) f  +1 ]C^{(+)}_{r}
\ee
The non-linear part of this Lagrangian is studied in Appendix \ref{nl} where we argue that in Schwarzschild coordinates there are no time derivatives acting on the ghosts in $l=1$ even sector.

\subsection {Covariant Quantization of the $l = 1$ Odd (Magnetic Dipole) Modes}\label{l1odd}
In the $l = 1$ odd (magnetic dipole) sector, there is a single gauge symmetry $\xi^{1m(-)}$ for each $m=-1,0,1$, which we fix using the 2D background covariant gauge-fixing condition 
\begin{equation}
	r^ah_a^{(-)}=0
\end{equation}
From \cref{gt}, we see that the gauge transformation \cref{gtPrime} of the gauge-fixing condition is
\be
 r^a \delta h_{a}^{(-)} =r^a  \mathcal{D}_a\xi^{(-)}  + r^af^{ (-)}_{a}[h,\xi]
	\label{haFreedomDipole}
\ee
This variation involves, at the quadratic level defining the ghost kinetic term, a derivative of 
$\xi^{(-)}$. This means that in general, these ghosts are propagating and coupled to gravitational modes. 
 The corresponding ghost action is
\be
 \int \bar C^{ (-)} \Big (r^a (\partial_a + V_a(h))\Big ) C^{(-) }
\label{mag}\ee
where the  coupling $r^aV_a(h)$ represents  vertices where the  ghosts for the $l=1$ odd modes interact with $h_{\mu\nu}$.

In Schwarzschild coordinates  where $r^a= (0,f)$ the ghosts action simplifies to
\be
 \bar C^{ (-)}  f (\partial_r + V_r(h)) C^{(-) }
\label{Sh}\ee
Here the propagator is instantaneous since the kinetic terms has only $r$-derivatives. There are no derivatives acting on this ghost field at higher order in $h_{\mu\nu}$, so in Schwarzschild coordinates there are no time derivatives acting on the ghosts at all in the $l = 1$ odd sector. 

\

\subsection {Covariant Quantization of the $l = 0$ (Monopole) Modes: Results to All Orders}\label{l0}
In the $l = 0$ (monopole) sector there is a single 2D vector $\xi_a^{00(+)}$, a parameter of a gauge symmetry, which we fix using the 2D background covariant gauge-fixing conditions 
\be
K= t^a r^b h_{ab}=0
\ee
The first one $K = 0$ is a 2D scalar while the second one $t^a r^b h_{ab} = 0$ is a 2D pseudoscalar since $t^{a} = -\epsilon^{ab}r_{b}$ is a pseudovector.

The gauge transformation \cref{gtPrime} of the gauge-fixing conditions is now given by
\begin{align}
	t^a r^b \delta  h_{ab} &= t^a r^b (\mathcal{D}_{a}\xi_{b}+ \mathcal{D}_{b}\xi_{a} + f_{ab}[h,K,\xi])
	\label{habFreedomMonopole}\\
	\delta K &= \frac{2}{r}\,r^{a}\xi_{a} + f_{(K)}[h,K,\xi]
	\label{KFreedomMonopole}
\end{align}
which leads to a ghost Lagrangian 
\be
\bar C^{(K)} \Big (\frac{2}{r}\,r^{a}C_{a} + f_{(K)}[h,C]
\Big ) + \bar C^{(h)} \Big ( t^a r^b (\mathcal{D}_{a}C_{b}+ \mathcal{D}_{b}C_{a} + f_{ab}[h, C])
\Big )
\ee
We can integrate over $\bar C^{(K)}$ which imposes the constraint
\be
\frac{2}{r}\,r^{a}C_{a} + f_{(K)}[h,C] = 0
\label{barK}\ee
The first term in the ghost action now vanishes. We now note that, since $f_{(K)}[h,C]$ is a Lorentz scalar, it can only depend on  $r^aC_a$ (not the pseudo-scalar $t^aC_a$), so that 
\be
f_{(K)}[h,C] = \hat f_{(K)}[h]\,r^aC_a
\label{leq0Constraint}
\ee
for some 
functional $\hat{f}_{(K)}[h]$ of $h_{ab}^{00}$. This means that the constraint \cref{leq0Constraint} really reads
\be 
\quad \Big (\frac{2}{r} + \hat f_{(K)}[h] \Big ) r^aC_a=0
\label{barK1}\ee
and so sets $r^aC_a = 0$  up to a local functional determinant. Such changes of variables were discussed in the Introduction below eq. \rf{gh} and simply contribute a term proportional to $\delta^2(0)$ to the local measure of integration. We neglect these following the arguments in \cite{Fradkin:1977hw}.

The second term in the Lagrangian is
\be
 \bar C^{(h)} \Big ( t^a r^b (\mathcal{D}_{a}C_{b}+ \mathcal{D}_{b}C_{a} + f_{ab}[h, C])
 \label{SecondTerm}
\Big )
\ee
In general the kinetic term, coming from the part of the action quadratic in ghosts, has both space and time derivatives, and a coupling is present between the ghosts and the 2D monopole sector metric $h^{00}_{ab}$, encoded in the term $\bar C^{(h)}  t^ar^bf_{ab}[h, C]$.

However, in the  Schwarzschild coordinates, we find from \rf{barK1} that $C_r=0$ and \cref{SecondTerm} reduces to
\be
 \bar C^{(h)} f (r)\Big [ \Big ( \mathcal{D}_{r} - {f'\over f} \Big )C_{t} + f_{tr}[h, C_t] \Big]
\ee
The propagator is again instantaneous since the kinetic terms has only $r$-derivatives, and the non-linear term $f_{tr}[h, C_t]$ has no derivatives acting on ghosts. Therefore there are again no time derivatives acting on the ghost fields in Schwarzschild coordinates.

\

\noindent {\textit{\textbf{Explicit All Orders in $h_{ab}^{00}$ Ghost Action}:}}  

\

The monopole $l=0$ sector has the advantage that the relevant parts decouple from other modes even at the non-linear level. Therefore we are able to find an explicit expression for the ghost action, including all non-linear terms.  With our gauge choices $K = t^ar^bh_{ab}^{00} = 0$ a convenient parameterization for the remaining two components of $h_{ab}$ is
\begin{equation}
	h_{ab}^{00} \equiv \frac{1}{f(r)^2}\,At_at_b + Br_ar_b
	\label{MonopoleGaugeabSpherical}
\end{equation}

The Lagrangian for the monopole ghosts is  given exactly (to all orders in $h_{\mu\nu}$) by
\begin{multline}
	\bar{C}^{(K)}\,\inp{\frac{2}{r}\frac{1}{1 + f(r)B}}r^a C_a 
	\\+ \bar{C}^{(h)}\insb{\inp{t^a\partial_a-\frac{f(r)(t^b\partial_bB)}{1 + f(r)B}}r^cC_c + \inp{r^a\partial_a-\frac{f(r)f'(r)-(r^d\partial_dA)}{f(r)-A}}t^c C_c}
	\label{MonopoleExplicit}
\end{multline}
Note that here the anti-ghost field $ \bar{C}^{(h)}$ is a 2D pseudo-scalar, same as the corresponding auxiliary field $B$ in $ Bt^ar^bh_{ab}$ in Sec. \ref{BRST} so the the total ghost action is a scalar since $ \bar{C}^{(h)}$ multiplies a piece of the Lagrangian which is linear in $t^a$.

Now we take into account equation  \rf{barK1}  and we find that the remaining ghost action is given by
\be
	 \bar{C}^{(h)} \Big (r^a\partial_a-\frac{f(r)f'(r)-(r^d\partial_dA)}{f(r)-A}\Big ) t^c C_c
\ee
In general this describes propagating ghosts coupled to monopole gravitational perturbations. However, 
in Schwarzschild coordinates we find only space derivatives acting on ghosts.
\be
	 \bar{C}^{(h)} f(r)\inp{\partial_r-\frac{f'(r)-\partial_rh^{00}_{tt}}{f(r)-h^{00}_{tt}}}  C_t
\ee
Therefore the propagator is instantaneous as in the cases above.

\section{Faddeev's Theorem and a Possible Hamiltonian Origin} \label{H}

\subsection{Gauges with Unitary and Pseudo-Unitary Hamiltonians}
\label{unitary}
In the Hamiltonian formalism in gauge theories where we have 1st class constraints $\phi^\alpha (t, \vec x)$ and additional conditions $\chi_\alpha (t, \vec x)$, the Poisson bracket 
\be
\{ \chi_\alpha (t, \vec x), \phi^\beta(t, \vec y) \} = M_\alpha {}^\beta \delta^3(\vec x- \vec y)
\label{pb}
\ee
defines a differential operator $M_\alpha {}^\beta$. 
As we suggested in the Introduction, there are two different classes of gauge-fixing functions functions $\chi_\alpha$ which either lead to a theory with a unitary Hamiltonian, or to a theory with the pseudo-unitary Hamiltonian.

\

\centerline {\it Case of Unitary Hamiltonian $H(p^*, q^*)$ in Gauge Theories}

\

The first class of conditions $\chi_\alpha(p,q)$ is usually  associated with Lorentz non-covariant gauges, like Coulomb gauge in Yang-Mills theory or Dirac gauge in gravity.  This class includes all instances where the {\it differential operator}  $M_\alpha {}^\beta$ {\it does not involve time derivatives}. In such a case there exists a unitary Hamiltonian in a space with $(n-m)$ physical degrees of freedom  $(p^*, q^{* })$, where $*=1, \dots, (n-m)$ and {\it which is manifestly ghost-free}. 

In covariant quantization as defined in \cite{Faddeev:1969su} one can arrange that 
\be
\delta \chi_\alpha =- \{ \chi_\alpha, \phi^\beta\} \xi_\beta
\label{delta} \ee
at $\chi_\alpha= \phi^\alpha=0$ 
Therefore the differential operator $M_\alpha {}^\beta$ defined by the Poisson bracket eq. \rf{pb} of canonical quantization actually defines the FP ghost action $\bar C^\alpha  Q_{\alpha}{}^{\beta} C_\beta$ in covariant quantization. When the operator $M_\alpha {}^\beta$ has only space derivatives one finds that, though a nontrivial FP ghost action is present in the covariant quantization, it has the particular feature that {\it the ghosts propagators are instantaneous}. 

It is known\footnote{We are grateful to A. Weinstein and M. Shifman for the clarifying discussion of this issue, namely the absence of ghosts in the canonical Feynman rules in the Coulomb gauge in Yang-Mills theory \cite{Khriplovich:1969aa}, \cite{Faddeev:1969su,Fradkin:1970pn}, and the presence of instantaneous ghosts as well as instantaneous gluons in the covariant Feynman rules. }  in the example of Coulomb gauge in Yang-Mills theory that the ghosts loops with instantaneous propagators are cancelled by the loops of the instantaneous part of the gluon part of the propagator, to all orders in perturbation theory \cite{Zwanziger:1998yf}. It is also known  that there are no closed instantaneous loops when the Feynman rules are deduced from the Hamiltonian path integral and the $S$ matrix is computed as time ordered product  of the unitary Hamiltonian.  
In this case the equivalence of the  Hamiltonian perturbative Feynman rules  in QCD \cite{Khriplovich:1969aa}, \cite{Faddeev:1969su,Fradkin:1970pn} and the Lagrangian De Witt-Faddeev-Popov rules in QCD is clearly established. In gravity in the Dirac gauge  \cite{Dirac:1958jc} the ghosts in the covariant quantization also have instantaneous propagators whereas the underlying Hamiltonian is unitary and the Hilbert space of physical degrees of freedom is ghost-free \cite{Faddeev:1969su,Fradkin:1970pn}. 

In Appendix  \ref{can} we explain the general relation between the original constrained canonical variables $q$ and $p$ in eq.  \rf{action} and the independent  physical canonical variables $(q^*, p^*)$ eq. \rf{fad}.

\

\centerline {\it Case of Pseudo-Unitary Hamiltonian in Gauge Theories,  $H(q^A, p_A,\eta^a, {\cal P}_a)$}

\

The other class involves situations where the condition $\chi_\alpha = \chi_\alpha (q,p, \lambda, \dot \lambda)$ depends not only on the na\"ive canonical variables $(q,p)$, but also on Lagrange multipliers $\lambda$ and their time derivative $\dot{\lambda}$, where the Lagrange multipliers originate from the imposition of the constraints in the Hamiltonian form of the action:
\be
S(q,p, \lambda) = \int \mathrm{d}t \inp{p_i \dot q^i -H(q,p) - \lambda _\alpha \phi^\alpha (q,p)}
\ee
For example, in QCD, the field $A_0$ is a Lagrange multiplier since the classical action does not depend on $\partial _0 A^0$. In the Lorentz covariant gauge $\partial _\mu A^\mu=0$, there is a time derivative of the  Lagrange multiplier $\lambda=A^0$, since $\partial _0 A^0=\dot \lambda$

The set of canonical coordinates $(q,p)$ is now enlarged by the Lagrange multipliers and their canonical momenta so that the total na\"ive phase space is spanned by $(q^A, p_A)$  with  $A=1, \dots , n+m$. This allows for the accommodation of Lorentz covariant gauges, like Feynman gauge in Yang-Mills theory and de Donder type gauges in gravity. In this case, one must add to the system $2m$ additional degrees of freedom $(\eta^a, {\cal P}_a)$, $a=1, \dots, 2m$, with opposite statistics, which correspond to propagating FP ghosts and anti-ghosts. The net number of commuting  minus anti-commuting canonical variables is $n+m-2m=n-m$, in agreement with the counting of physical states $(q^*,p^*)$ in the unitary gauges described above, where neither Lagrange multipliers nor FP ghosts/antighosts are propagating degrees of freedom. The Hamiltonian in such an extended space $(q^A, p_A,\eta^a, {\cal P}_a)$ defines an S-matrix which is pseudo-unitary 
in a state space with indefinite metric, as explained in \cite{Fradkin:1977hw}.

Examples of  gauges with pseudo-unitary Hamiltonian  in gravity include harmonic/de Donder gauge, which has propagating (non-instantaneous) FP ghosts in the covariant quantization method. The underlying Hamiltonian in this class of gauges is pseudo-unitary. The proof of unitarity of the S-matrix in this class of gauges, where ghost degrees of freedom appear even in the canonical/Hamiltonian construction, follows only because of the equivalence of this S-matrix to the one in the class of gauges with the ghost-free unitary Hamiltonian.

\subsection {$l\geq 2$}
Our 4 gauge fixing functions, for each $(l,m)$ with $l \geq 2$, are   $\chi_\alpha = \{ j_{a}, G, h_2\}$. To leading order in $h_{\mu\nu}$ they transform under the 4 gauge parameters $\xi_\alpha = \{ \xi_a, \xi^{(+)}, \xi^{(-)} \}$ with the same $l$ and $m$ via
\be
\delta \chi_\alpha =- \{ \chi_\alpha, \phi^\beta\} \xi_\beta + O(h_{\mu\nu})
\label{deltaChi}
\ee
where the first term on the right hand side of \cref{deltaChi} is given by the leading order terms in eqs. \rf{jaFreedomgeq2}-\rf{h2Freedom}. The corresponding leading order ghost action---in particular the ghost kinetic term---is associated with the determinant of the matrix
$\det ||\{ \chi_\alpha, \phi^\beta\} || \neq 0$ as explained in  \rf{det}, which is given by 

	\be
	\det\qmatrix{\mathds{1} & \mathcal{D}_a-\frac{2}{r}r_a & 0\\0 & \frac{2}{r^2} &0\\0 & 0 & 2} = \det\qmatrix{\mathds{1} & 0 & 0\\0 & \frac{2}{r^2} &0\\0 & 0 & 2}
	\ee
	Such a determinant is algebraic and will contribute to the action as $\delta^2(0)$. We can see therefore that in both covariant as well as in canonical quantization there are not expected to be propagating ghosts for the modes with $l\geq 2$. One can further show that in Schwarzschild coordinates, time derivatives of ghosts are absent even in the 
higher order terms. Thus, the absence of a nontrivial ghost action for the $l\geq 2$ modes provides evidence that a unitary Hamiltonian is available for these modes.

\subsection {$l=1$  even}
Our 3 gauge fixing functions, for each $m=-1,0,1$, are $\chi_\alpha = \{ j_{a}, K \}$. 
To leading order they transform under the 3 gauge parameters $\xi_\alpha = \{ \xi_a, \xi^{(+)}  \}$ with the same $m$ via the Poisson bracket $\delta \chi_\alpha =- \{ \chi_\alpha, \phi^\beta\} \xi_\beta + O(h_{\mu\nu})$ given by the leading order terms in eqs. \rf{jaFreedom1} and \rf{KFreedom}. The corresponding 
	determinant $\det||\{ \chi_\alpha, \phi^\beta\} ||$ is given by
	\be
	\det\qmatrix{1 & \mathcal{D}_a-\frac{2}{r}\,r_a\\\frac{2}{r}\,r^a & -\frac{2}{r^2}} = \det\inp{-\frac{2}{r^2}}-\det\inp{\frac{2}{r}\,r^a\inp{\mathcal{D}_a-\frac{2}{r}\,r_a}}\ee
In Schwarzschild coordinates, with $r_a= (0,1)$ and $r^a= (0, f)$, this becomes
	\be
	\left.\det\qmatrix{1 & \mathcal{D}_a-\frac{2}{r}\,r_a\\\frac{2}{r}\,r^a & -\frac{2}{r^2}}\right|_{\mathrm{Schld}} =  \det\inp{-\frac{2}{r^2}}-\det\inp{\frac{2f(r)}{r}\inp{\mathcal{D}_r-\frac{2}{r}}}
	\ee
	
Note that this determinant has  only  space derivatives. We can deduce therefore from the  canonical quantization in Schwarzschild coordinates, that the FP ghosts in a covariant quantization will have instantaneous propagator. One can further show that in Schwarzschild coordinates, time derivatives of ghosts are absent even in the higher order terms. Again, the absence of the time derivatives in the ghosts action for $l= 1$ even modes provides an evidence that the unitary Hamiltonian is available for these modes.
	
	\subsection{$l=1$  odd}
	Our 1 gauge fixing function, for each $m=-1,0,1$, is $\chi_\alpha = \{r^a h_a \}$ which to leading order transforms under a the gauge parameter $\xi_\alpha = \{  \xi^{(-)}  \}$ with the same $m$ via the Poisson bracket $\delta \chi_\alpha =- \{ \chi_\alpha, \phi^\beta\} \xi_\beta + O(h_{\mu\nu})$ given by the leading order terms in eq. \rf{haFreedomDipole}.
	In Schwarschild coordinates we see the corresponding determinant is given by $\det||\{ \chi_\alpha, \phi^\beta\}|| = \det(\mathcal{D}_r)$, which involves a derivative operator in the $r$ direction only. The corresponding FP ghost action shown in eq. \rf{mag} is not vanishing but leads to ghost loop diagrams with an instantaneous propagator. 
	
	Thus, the ghost action for magnetic dipoles, which has only space derivatives in the action in Schwarschild coordinates, provides evidence that the unitary Hamiltonian is available for these modes.

\subsection{$l=0$}
In Schwarschild coordinates we have the two gauge-fixing conditions $\chi_\alpha = \{K, h_{tr}\}$ and the two gauge parameters $\xi_{r}, \xi_t$. The determinant of the corresponding Poisson bracket
$\delta \chi_\alpha =- \{ \chi_\alpha, \phi^\beta\} \xi_\beta$ is given to all orders in $h_{ab}^{00}$ by
\begin{multline}
	\det \left(\begin{array}{cc}{2\over r} {f(r) \over 1+ f(r) h_{rr} } & \, \, \, 0 \\
		\cr
		f(r)\inp{\partial_t- {f(r)  \partial_t  h_{rr} \over 1+ f(r) h_{rr} }} & \hskip 0.5 cm  \,  \, f(r)\inp{\partial_r -  {f'(r) - \partial_r  h_{tt} \over f(r) -h_{tt}}}\end{array}\right)
	\\= \det \left(\begin{array}{cc}{2\over r} {f(r) \over 1+ f(r) h_{rr} } & \, \, \, 0 \\
		\cr
		0 & \hskip 0.5 cm  \,  \, f(r)\inp{\partial_r -  {f'(r) - \partial_r  h_{tt} \over f(r) -h_{tt}}}\end{array}\right)
\end{multline}

Here again we can see that the determinant of the Poisson bracket is in agreement with the expression for the action of the FP ghosts in the covariant method. There are only space derivatives on monopole FP ghosts. Therefore in the covariant quantization they have instantaneous propagators and it is expected that their loops will cancel with the instantaneous part of the gravitational fields.

Time derivatives in Schwarzschild coordinates explicitly vanish to all order here. Thus, the ghosts action for monopoles, which has only space derivatives in the action, provides evidence that the unitary Hamiltonian is available also in the monopole sector.

\section{Discussion}
A rather surprising feature of the covariant quantization of perturbative gravity in the Schwarzschild black hole background 
in Regge-Wheeler gauge is that FP ghosts are absent (i.e. decoupled) for all modes with the exception of monopoles and dipoles. These latter modes are known to be related to the ADM mass and angular momentum of the perturbed black hole; for example, on-shell and to linear order, they are known to encode the linearized perturbation of the ADM mass and the linearized rotation of the perturbed black hole solution in presence of sources \cite{Regge:1957td,Zerilli:1971wd,Martel:2005ir}. 

In the well known background-covariant and Lorentz-covariant gauges in \cite{tHooft:1974toh,Goroff:1985th,vandeVen:1991gw} the gauge fixing functions---e.g. for de Donder gauge $F_\mu=\nabla_\nu^{(g )} h^\nu{} _\mu -{1\over 2} \nabla_\mu^{(g )} h^\nu{ }_\nu$---depend on background  covariant derivatives of the gravitational perturbation $h_{\mu\nu}$.The ghosts are propagating due to the tensorial nature of the gravitational perturbations 
$\delta h_{\mu\nu} = 2\overline{\nabla}^{(\bar{g})}_{(\mu } \xi^{}_{\nu)}$ which involves  both space and time derivatives acting on $\xi_\nu$. 
In gravity, Dirac gauge \cite{Dirac:1958jc} is the only one known to have FP ghosts with instantaneous propagators in a covariant quantization, as shown in \cite{Faddeev:1969su,Fradkin:1970pn,Faddeev:1973zb}, and  to have a unitary Hamiltonian (in the sense described in \cref{unitary} above).

Here we have studied  covariant quantization of perturbative gravity in a Schwarzschild back hole background, using the Regge-Wheeler \cite{Regge:1957td,Zerilli:1971wd,Martel:2005ir} framework where the quantum fields are expanded in spherical harmonics. In the Schwarzschild black hole background there is a natural split of the space into a warped product $\mathcal{M} = \mathcal{M}_2\times \S^2$  of two 2D submanifolds. All dependence on the $\S^2$ 
is encoded in the the discrete indices of the harmonics, $l, m$ and $\pm$. For each $(l,m,\pm)$, the quantization is reduced to a quantization of a quasi-two dimensional theory on $\mathcal{M}_2$ (with the additional information of a radial function $r(x)$): The gravity perturbations with fixed values of $(l,m,\pm)$ depend only on $x^a, a=1,2$, the two coordinates of $\mathcal{M}_2$.  For low multipoles for which the Regge-Wheeler gauge condition, is not valid, we have proposed a two-dimensionally background covariant set of gauges in eq. \rf{ourG}. They can can written out using any choice of coordinates, including, for example, Schwarzschild, Eddington-Finkelstein or Kruskal-Szekeres coordinates. We found the following:

\

1. For all even and all odd modes with $l\geq 2$, the corresponding FP ghosts are not propagating: their kinetic terms do not contain derivatives.

2. All even and all odd dipole modes with $l=1$, and monopole modes with $l=0$, have nontrivial FP ghosts in covariant quantization, since their kinetic terms involve derivatives. 

\

\noindent In the special case of Schwarzschild coordinates our results are: 

\

1. In Schwarzschild coordinates, the low multipole 
ghosts have no time derivatives, only space derivatives. 
Their propagators are therefore
instantaneous, as in 
Coulomb gauge in QCD \cite{Khriplovich:1969aa,Faddeev:1969su,Fradkin:1970pn} and 
Dirac gauge in gravity  \cite{Faddeev:1969su,Fradkin:1970pn,Faddeev:1973zb}. 

2. We have provided evidence that with our choice of gauge-fixing functions our covariant quantization rules  
when viewed in Schwarzschild coordinates
are consistent with the existence of 
an underlying unitary Hamiltonian in a manifestly ghost-free Hilbert space with a positive-definite metric, 
as suggested by Faddeev's theorem \cite{Faddeev:1969su}.

\

An open issue which needs to be addressed with regard to a potential canonical quantization of gravity in the black hole background concerns the fact that the existing constructions of the canonical (Hamiltonian) path integral in \cite{Faddeev:1969su,Fradkin:1970pn,Faddeev:1973zb,Fradkin:1976xa,Fradkin:1977hw} were only performed in a flat Minkowski background. Meanwhile in the black hole background in Schwarzschild coordinates there is an event horizon along which the relevant coordinates break down. The concept of the Hamiltonian and of the 
Hilbert space of states might be more  subtle, if well-defined at all.

Nevertheless, 
Faddeev's theorem  \cite{Faddeev:1969su}, valid at least in the flat Minkowski background, suggests that a ghost-free unitary Hamiltonian ought to exist  
for the gauge-fixing conditions studied here in the black hole background in Schwarzschild coordinates. Namely, we have found that in the covariant quantization in Schwarzschild coordinates, there are no time derivatives acting on the ghost fields. This suggests that  such a Hamiltonian, if  explicitly constructed, might belong to the class we described in \cref{unitary}, the case of the unitary ghost-free Hamiltonian $ H(p^*, q^{* })$ depending on $(n-m)$ degrees of freedom  described by $2(n-m)$ canonical variables $(p^*, q^{* })$.

The reason this is a likely outcome of the canonical quantization is that the case of pseudo-unitary Hamiltonian in a Hilbert space of states with the indefinite metric, also described in \cref{unitary}, $H(q^A, p_A,\eta^a, {\cal P}_a)$, would be inconsistent with the absence of time derivatives acting on the ghosts, which we found in this paper. Here $A=1, \dots , n+m$ involves commuting fields, and  $a = 1, \dots , 2m$ involves anti-commuting fields. The total counting of degrees of freedom is therefore $n+m-2m=n-m$. But we have just shown that all of our anti-commuting fields (FP ghosts and anti-ghosts)  have no time derivatives, so they are not expected to contribute to a 
space of states with indefinite metric in a process of canonical quantization. 

Note that in Eddington-Finkelstein and  Kruskal-Szekeres coordinates the situation is different and has to be studied separately. Although we have performed a covariant quantization which is valid in any of these coordinate systems, the canonical quantization is still to be explored. 

To summarize, it would be very interesting to perform a canonical quantization of gravity in the Schwarzschild black hole background in the class of gauges presented here. We leave this for future work.

\section*{Acknowledgement}
We are grateful to  A. Barvinsky, R. Bond, V. Chandrasekaran, R. Flauger, A. Linde, R. Mahajan, M. Mirbabayi, E. Poisson,  G. Satishchandran, M. Shifman, V. Shyam, E. Silverstein, P. Stamp, D. Stanford, A. Starobinsky, A. Vainshtein, A. Van Proeyen, I. Volovich,   and R. Wald  for stimulating and helpful discussions.  
RK and AR are supported by the Stanford Institute of Theoretical Physics and by the Grant PHY-2014215 of National Science Foundation
 (United States/US). RK is additionally supported by the  Simons Foundation Origins of the Universe program (Modern Inflationary Cosmology collaboration (United States/US). AR is additionally supported by the National Science Foundation GRF Program under Grant No. DGE-1656518 and by a Fletcher Jones Foundation National Science Foundation Graduate Fellowship in the Stanford School of Humanities \& Sciences (United States/US).
 
\appendix

\section{De Witt-Faddeev-Popov Covariant Quantization of Gravity}
\label{cov}
We consider 4D asymptotically flat Einstein gravity in the background of a Schwarzschild black hole $(\mathcal{M}, g_{\mu\nu})$. The total classical action depends on $\bar g_{\mu\nu}=g_{\mu\nu}+h_{\mu\nu}$, where $h_{\mu\nu}$ is the quantized perturbative gravitational field (graviton field) and the background field is the Schwarzschild black hole metric $g_{\mu\nu}$. The action $S(g+h)$ is invariant under the  gauge symmetries
\begin{equation}
	\delta h_{\mu\nu} = \overline{\nabla}^{(g +h)}_{\mu} \xi_{\nu} + \overline{\nabla}^{(g +h)}_{\nu} \xi_{\mu}, \qquad \delta g_{\mu\nu} = 0
	\label{SchldGaugeInv}
\end{equation}
where $\overline{\nabla}^{(g +h)}_{\mu}$ is the covariant derivative operator of the full metric $g+h$. 

Due to the gauge symmetries the naive path integral 
\be
 \int  D h
 \, e^{\mathrm{i} S ( g +h ) } 
\ee
has to be defined using De Witt-Faddeev-Popov procedure \cite{DeWitt:1967ub,Faddeev:1967fc}. This procedure in the simple case suitable for our purpose involves a set of gauge-fixing conditions $\chi_\alpha(g;h)=0$ which constrain the gravitational fields. The path integral becomes
\be
 \int  D h
 \,J_\chi ( g;h)  \,\delta \big (\chi_\alpha ( g;h)\big )\, e^{\mathrm{i} S ( g+h) } 
\label{pi}\ee
Here the Jacobian $J_\chi (g; h)$ is defined by the variation of the gauge-fixing function $\chi_\alpha(g; h)$ under the gauge symmetry with the parameters $\xi_\beta$
\be
J_\chi (g, h)  = \exp {\Tr \ln } \, Q_{\alpha}{}^{ \beta} (g, h) 
\ee
where
\be
\delta \chi_\alpha = Q _{\alpha}{}^{ \beta} (g, h) \,\xi_\beta
\ee
This Jacobian  can be also presented with the help of the FP ghosts \cite{Faddeev:1967fc} as follows
\be
J_\chi = \int D\bar C^\alpha DC_\beta \, e^{\mathrm{i} \int \mathrm{d}^4 x \, \bar C^\alpha(x)  Q_{\alpha} {}^{ \beta}  (g, h)C_\beta(x) }
\ee

When $Q_{\alpha} {}^{ \beta} (g;h)$ is a local function of $(g;h)$ without differential operators acting on $C_\beta(x)$, the relevant ghost action becomes $\bar C^\alpha(x)  \tilde C_\alpha(x)$ where  $\tilde C_\alpha(x) = Q_{\alpha} {}^{ \beta}  (g, h)C_\beta(x)$. The corresponding ghosts are non-propagating and drop from Feynman rules. When $Q_{\alpha} {}^{ \beta}   (g;h)$ involves   differential operators with time and space derivatives, the corresponding ghosts are propagating and generically give important contribution to the Feynman diagrams. When $Q_{\alpha} {}^{ \beta}  (g;h)$ involves  a differential operator with only space derivatives, the corresponding ghosts are said to have ``instantaneous" propagators.

\section{Faddeev-Fradkin-Vilkovisky Canonical Quantization of Gravity}\label{can}

In this appendix, we would like to  present a brief summary of the most relevant results obtained in \cite{Faddeev:1969su,Faddeev:1973zb,Fradkin:1976xa,Fradkin:1977hw} concerning the canonical quantization of gravity. We would like to stress that the standard Faddeev-Fradkin-Vilkovisky  quantization of gravity was formulated in the context of a flat Minkowski background, whereas in this paper we  perform the quantization in the Schwarzschild black hole background using the covariant quantization method developed by De Witt \cite{DeWitt:1967ub,tHooft:1974toh,Kallosh:1974yh,Grisaru:1975ei}.

It might be useful to clarify the relevant observation made by Weinberg in  Ch. 15.8 on p. 41 in \cite{Weinberg:1996kr}. He noticed that in ``theories like general relativity there is no way of choosing a coordinate system in which the ghosts decouple. Such theories may be dealt with by the BRST-quantization method described at the end of the previous section, using BRST invariance to prove that the S-matrix in a physical ghost-free Hilbert space is unitary." 

To clarify this statement we need to explain the precise  meaning of the words ``the ghosts decouple": one has to make a clear distinction between ghosts decoupling in a covariant BRST quantization procedure and in canonical quantization. For example, one may wonder what exactly the statement above means for e.g. Dirac gauge 
\cite{Dirac:1958jc} in view of the fact that in \cite{Faddeev:1969su,Fradkin:1970pn,Faddeev:1973zb} gravity was canonically and covariantly quantized in the Dirac gauge, with no FP ghosts appearing in the canonical quantization. The clarified statement is: in canonical quantization in Dirac gauge in gravity there is a unitary Hamiltonian and there are no FP ghosts. In covariant quantization in the Dirac gauge, using the BRST method, there are FP ghosts with instantaneous propagators. To explain this we proceed with a review of the results in  \cite{Faddeev:1969su,Fradkin:1970pn,Faddeev:1973zb}.

The Faddeev-Fradkin-Vilkovisky construction involves defining the canonical (Hamiltonian) Feynman integral for systems with {\it singular Lagrangians}, which have the property that the equation
\be
p_i = {\partial L(q, \dot q)\over \partial \dot q^i}
\ee
cannot be solved for $\dot q^i$ as a function of the $q^i$'s and $p_i$'s, which is a constrained system studied first by Dirac. The Hamiltonian $H(q^i, p_i), i= 1, \dots, n$ depends on  2n variables. For such singular Lagrangians with $m$ gauge symmetries, the na\"ive canonical variables  $(q^i, p_i)$ of the na\"ive phase space $\Gamma$ satisfy a set of first class constraints 
\be
\phi^\alpha (q,p) =0, \, \qquad \alpha=1,\dots m
\label{contraints}
\ee
These are in involution with each other as well as with the Hamiltonian. The constraints  define a surface $M$ of dimension $(2n-m)$ in $\Gamma$.

For such Lagrangians $L(q, \dot q)$ all equations of motion, including the constraints, can be obtained from the constrained action
\be
S(q,p, \lambda) = \int \mathrm{d}t \inp{p_i \dot q^i -H(q,p) - \lambda _\alpha \phi^\alpha (q,p)}
\label{action}\ee
For such systems the Feynman path integral is not well-defined unless an additional set of $m$ conditions $\chi_{\alpha}(q,p)$ on the canonical variables is introduced. These conditions\footnote{A more general choice of conditions is $\chi_a (q,p, \lambda, \dot \lambda)$ where there is a dependence on the Lagrange multiplier and its time derivatives. In these cases the underlying Hamiltonian was also constructed, and depends on $(n+m)$ degrees of freedom. The S-matrix in this case is pseudo-unitary, since the Hilbert space has an indefinite metric \cite{Fradkin:1977hw}. This case includes Lorentz covariant gauges like a harmonic gauge.}
\be
\chi_\alpha (q,p)=0
\label{nonr}
\ee
define a submanifold $\Gamma^*$ in $M$. These functions must satisfy the condition that 
\be
\det ||\{ \chi_\alpha, \phi^\beta\} || \neq 0
\label{det}\ee
since only in this case can the surface $\Gamma^*$ be defined.  Here $\{f,g\}$ are Poisson brackets in $\Gamma$. Here we remind the reader that, as explained near eq. \rf{delta}, in covariant quantization one can arrange that at $\chi_\alpha= \phi^\alpha=0$ the variation of the function $\chi_\alpha$ is
\be
\delta \chi_\alpha =- \{ \chi_\alpha, \phi^\beta\} \xi_\beta
\ee 
Therefore the Poisson bracket 
\be
\{ \chi_\alpha (t, \vec x), \phi^\beta(t, \vec y) \} = M_\alpha {}^\beta \delta^3(\vec x- \vec y)
\ee
of the canonical quantization defines (via $M_\alpha {}^\beta$) the FP ghost action $\bar C^\alpha  Q_{\alpha}{}^{\beta} C_\beta$ in covariant quantization. 

A convenient choice is when $\{\chi_\alpha, \chi_\beta\}=0$, in which case we can perform a canonical transformation in $\Gamma$ to obtain the new canonical variables
\be
p'_\alpha =\chi_\alpha (p,q) =0\, ,  \qquad q^{'\alpha} = q^{'\alpha} (p^*, q^*)
\ee
where $q^{'\alpha}$ and $p'_\alpha$ are canonically conjugate, and where the $q^{'\alpha}$ are parameterized by the physical canonical variables $q^*, p^*$, which are independent coordinates on $\Gamma^*$. Therefore on  $\Gamma^*$ we are left with  $2(n-m)$ independent canonical variables $( p^*, q^*)$.
In such case the correct Feynman integral can be given in a canonical/Hamiltonian form as
\be
\int \prod_{t}   {\mathrm{d}p^* \mathrm{d}q^{*} \over 2\pi} \, \exp \inb{\mathrm{i} \insb{p^* \dot q^{*}- {\cal H} (p^*, q^*)}\mathrm{d}t}
\label{fad}\ee
This Hamiltonian defines the unitary S-matrix and depends on the $2(n-m)$ physical canonical variables, i. e. on $(n-m)$ physical degrees of freedom. The Hilbert space of physical states has a positive-definite metric and is manifestly ghost-free.

The basic result of Faddeev-Fradkin-Vilkovisky for the class of constraints in eq. \rf{nonr} is that the path integral in \rf{fad} based on a unitary Hamiltonian can be given in the form corresponding to a covariant quantization as defined in \cite{DeWitt:1967ub,Faddeev:1967fc}, where the constraint function $\chi_{\alpha}(q,p)$ defines a gauge-fixing function $\chi_{\alpha}(\phi)$, depending on the classical fields of the theory. In addition, observables computed using this path integral will be independent on the specific choice of the additional condition/gauge-fixing function.

A simple example of this situation is the Coulomb gauge in QCD. In this case there is a unitary Hamiltonian with only physical degrees of freedom. In notation of \cite{Faddeev:1969su} there is a  constraint with the Lagrange multiplier $A^0$. The configuration space coordinate is $q^i= A^i$ with canonical conjugate $p_k= E_k\equiv F_{0k}$. The constraint and additional condition are
\be
\phi (q,p) = \partial_k E_k +[A_k, E_k]\, , \qquad  \chi (q, p)= \partial_i A_i
\ee
The Poisson bracket  $\{ \chi_\alpha, \phi^\beta\}$ is computed in this case and it gives
\be
\{ \chi_\alpha, \phi^\beta\} = - [\delta^{\alpha \beta} \partial_i \partial_i + \epsilon^{\alpha \beta \gamma} A_k ^\gamma \partial_k ] \delta ^3 (\vec x-\vec y)
\ee
It is clear from this example why the determinant of the Poisson bracket at equal times $\{ \chi_\alpha(q,p) , \phi^\beta (q,p)\}$ can only produce space derivatives, and not time derivatives. 

The quantization of gravity in Dirac gauge performed in  \cite{Faddeev:1969su,Fradkin:1970pn,Faddeev:1973zb} has the same property: There is a unitary Hamiltonian with only physical degrees of freedom. In the covariant Lagrangian quantization there are ghosts, but only with instantaneous propagators.

\section {Non-Linear Part of the Ghost Action in Schwarschild Coordinates }\label{nl}
Here we will argue that the higher order in $h_{\mu\nu}$ part of the ghost action corresponding to the gauge-fixing conditions \rf{ourG} does not involve time derivatives acting on ghosts when written in Scwarzschild coordinates.

\subsection{$l \geq 2$ Modes}
We expect that the scalar  $f_{(G)}^{(+)}[h,C] $ in eq. \rf{GFreedom} can depend on the scalars $r^aC_a$ and $C^{(+)}$ (but not on the pseudoscalar $t^aC_a$) so that the constraint \cref{fG} can be expanded as
\be
\frac{2}{r^2}\,C^{(+)}  + \tilde f_{(G)}^{(+)}[h] \,C^{(+)} +  \hat f_{(G)}^{(+)}[h] \,r^b C_b =0
\ee
which allows us to express $C^{(+)}$ as a multiple (with an $h_{\mu\nu}$-dependent coefficient) of $r^aC_a$.
It follows that, after imposing this constraint, $f_{(G)}^{(+)}[h,C] $ depends only on $r^bC_b$
\be
f_{(G)}^{(+)}[h,C] \to -\hat F[h]\,r^b C_b
\ee

We similarly expect that the vector $f^{(+)}_a[h,C]$ appearing in eq. \rf{jaFreedomgeq2} depends on $C_a$ and on $r_ar^bC_b$ (since $C^{(+)}$ is now a function of $r^bC_b$), i.e. $f^{(+)}_{a}[h,C]= \alpha[h]\,C_a + \beta[h]\,r_a r^b C_b$, so that the residual ghost action \rf{l2} reads
\be
\bar C^a\big(1+\alpha[h]\big)C_{a} + \bar C^a\Big ({r^2\over 2} (\mathcal{D}_{a}  - \frac{2}{r}\,r_{a}) \hat F[h] r^b C_b + \beta[h]\, r_a r^b C_b\Big) 
\ee 
In  Schwarzschild coordinates ($t,r$) where $r_a =(0,1) , r^a= (0, f) $  there is a dependence on $C_t$ only in the first term
\be
\bar C^t\big(1+\alpha[h]\big)  C_{t}  
\ee 
so we can integrate over $C_t$ to impose the constraint $\bar C^t=0$, which reduces the ghost action to
\be
\bar C^r {r^2\over 2} \Big (\mathcal{D}_{r}+ \inp{\beta[h]-\tfrac{2}{r}} \Big )f(r)\,C_r 
\label{NL}\ee
We thus find that in Schwarzschild coordinates, the ghost action does not contain any time derivatives acting on the ghosts fields even at the higher order in $h_{\mu\nu}$ level. This indicates that the ghost free unitarizing Hamiltonian may be expected in this sector of the theory.

\subsection{$l = 1$ Even Modes} 
The argument is about the same as in $l\geq 2$ case. 
As argued above, we expect that the higher order in $h_{\mu\nu}$ terms $f[h,C]$ may depend on $\bar C^a$ contracted with $r_ar^bC_b$ or $C_{a}$. In Schwarzschild coordinates they therefore depend on $\bar{C}^rQ[h]C_r$, $\bar{C}^a\mathscr{D}_aQ[h]C_r$, or $\bar C^tQ[h]C_{t}$ where $Q[h]$ is some functional of $h_{\mu\nu}$ which does not contain differential operators acting on $C_t$ or $C_r$. We therefore expect that the higher order in $h_{\mu\nu}$ terms will not change the condition $\bar C^t=0$, which we derived earlier at leading order in $h_{\mu\nu}$. We therefore expect that terms with time derivatives acting on ghosts will not appear at higher orders in $h_{\mu\nu}$, since the only terms with derivatives will be $\bar{C}^a\mathcal{D}_aC_r \to \bar{C}^r\mathcal{D}_rC_r$. We therefore expect that the higher order in $h_{\mu\nu}$ ghost action for the $l = 1$ even modes will be given by an expression analogous to the one in eq. \rf{NL}, above, with all time derivatives acting on the ghosts dropping from this expression in Schwarzschild coordinates.

In  case of  $l =1$ odd modes  and  $l=0$ modes the absence of time derivatives in Schwarzschild coordinates to all orders was already demonstrated in Secs. \ref{l1odd} and \ref{l0}, respectively.

\section{Comments on Gravitational Propagator in the Black Hole Background }
In  \cite{Gaddam:2020rxb} 
an expression  for the tensorial part of the gravitational propagator in 
Regge-Wheeler gauge for the even modes was proposed 
as
 \begin{equation}
    \mathcal{P}^{abcd} ~ = ~ \dfrac{1}{4}\,f_\ell\left(k^2\right) \left(\eta^{ac}\eta^{bd} + \eta^{ad}\eta^{bc}\right) \, .
\label{Gerh}\end{equation}
where 
\begin{equation}\label{eqn:formfactor}
    f_{\ell}\left(k^2\right) =  - \dfrac{4 R^2_S}{\left(\lambda+1\right)} - \dfrac{2 R^4_S k^4}{\left(\lambda + 1\right)\left(\lambda - 3\right)\left(k^2 + R^{-2}_S \lambda\right)} \, ,
\end{equation}
and $\lambda \equiv  \ell^2 + \ell + 1 $.
The relevant part of the effective two-dimensional theory was taken in the form
\begin{equation}
\label{eqn:2dactionUU}
    S_{2d} ~ = ~ \sum_{l,m}\dfrac{1}{4}\int \mathrm{d}^2x \left(\mathfrak{h}^{ab}_{lm} \Delta^{-1}_{abcd}\mathfrak{h}^{cd}_{lm} + \mathfrak{h}^{ab}_{lm} \Delta^{-1}_{L,ab} \mathcal{K}_{lm} + \mathcal{K}_{lm} \Delta^{-1}_{R,ab} \mathfrak{h}^{ab}_{lm} + \mathcal{K}_{lm} \Delta^{-1} \mathcal{K}_{lm} \right) \, ,
\end{equation}
for all $l$. It was important here that $\mathcal{K}_{lm} \not = 0$. 

Meanwhile our results show that the action in cases $l=0,1$ has to be considered separately from $l\geq 2$ cases since {\it in these cases Regge-Wheeler gauge fixing with $\mathcal{K}_{lm} \not = 0$ is not valid}.   In particular, as we see from equations \rf{ourG}, for low multipoles  $l=0,1$  
 \be
\mathcal{K}_{lm}=0 \ , \qquad l=0,1 
\ee 
It is therefore not surprising that the propagator in \rf{Gerh}, \rf{eqn:formfactor} has strange properties, noticed in  \cite{Gaddam:2020rxb} :  the $l = 0$ mode causes a change in sign in the second term, the $l = 1$ mode has a pole.

 Thus, the  derivation of the propagator \rf{Gerh}, \rf{eqn:formfactor} from the action \rf{eqn:2dactionUU} is not valid for low multipoles. 
Instead of that, for $l=0,1$ one should consider the action \rf{eqn:2dactionUU} with $\mathcal{K}_{lm}=0$  
\begin{equation}
\label{eqn:2daction}
    S_{2d} ~ = ~ \sum_{l=0, 1; m}\dfrac{1}{4}\int \mathrm{d}^2x \  \mathfrak{h}^{ab}_{lm} \Delta^{-1}_{abcd}\mathfrak{h}^{cd}_{lm}   \, .
\end{equation}
Even this action is still a bit dangerous, since the gauge-fixing condition $t^a r^b h_{ab}=0$ still has to be added into account for the monopole ghosts.

We leave the derivation of the full graviton propagator in the Schwarzschild black hole background for future work.

\bibliographystyle{JHEP}
\bibliography{lindekalloshrefs}
\end{document}